\documentclass[11pt]{article} 
\usepackage[utf8]{inputenc}
\usepackage{multirow}
\usepackage{amsfonts}
\usepackage{amsmath}
\usepackage{amssymb}
\usepackage{amsthm}
\usepackage{caption}
\usepackage[dvipsnames]{xcolor}
    \definecolor{darkgreen}{rgb}{0,0.5,0}
    \definecolor{darkblue}{rgb}{0,0,0.6}
    \definecolor{purple}{rgb}{0.4,.2,0.7}
\usepackage[margin = 2.5cm]{geometry}
\usepackage{graphicx}
\usepackage[hyperfootnotes = false, colorlinks = true, linkcolor = darkblue, citecolor = purple]{hyperref}
\usepackage{subcaption}
\usepackage{ytableau}

\usepackage{csquotes}

\usepackage{tikz}
\usetikzlibrary{decorations.pathmorphing}
\usetikzlibrary{decorations.markings}
\definecolor{mathred}{RGB}{180,44,37}
\definecolor{mathblue}{RGB}{39,94,190}
\tikzset{>=latex} 
\tikzset{ photon/.style={decorate, decoration={snake}, draw=black}}

\usepackage{comment}

\usepackage{physics}

\newcommand{\be}{\begin{equation}}
\newcommand{\ee}{\end{equation}}
\newcommand{\bea}{\begin{eqnarray}}
\newcommand{\eea}{\end{eqnarray}}

\def\tr{\mathrm{tr}}

\begin{document}

\thispagestyle{empty}
\begin{center}
    ~\vspace{5mm}

  \vskip 2cm 
  
   {\LARGE \bf 
       Failure of the large-N expansion in a bosonic tensor model
   }

   \vspace{0.5in}
     
   {\bf Aidan Herderschee and Michael Winer
   }

    \vspace{0.5in}

  Institute for Advanced Study, Princeton, NJ 08540, USA
                
    \vspace{0.5in}

    \vspace{0.5in}

\end{center}

\vspace{0.5in}

\begin{abstract}
We study the tensor model generalization of the quantum $p$-spherical model in the large-$N$ limit. While the tensor model has the same large-$N$ expansion as the disordered quantum $p$-spherical model, its ground state is superextensive, in contradiction with large-$N$ perturbation theory. Therefore, the large-$N$ expansion of this model catastrophically fails at arbitrarily large-$N$, without any obvious signal in perturbation theory.

\end{abstract}

\vspace{1in}

\pagebreak

\setcounter{tocdepth}{3}
{\hypersetup{linkcolor=black}\tableofcontents}

\section{Introduction}\label{sec:introduction}

The large-$N$ expansion is a powerful computational tool for studying many-body systems. At the level of perturbation theory, the large-$N$ expansion corresponds to restricting to a small subclass of diagrams that are (usually) computationally tractable \cite{tHooft:1973alw}. However, the validity of the large-$N$ expansion is not well understood and is frequently assumed without explicit verification. In many models of interest, such as the Sachdev–Ye–Kitaev model \cite{Sachdev_1993,kitaev2015holography,Maldacena:2016hyu}, this assumption is valid. However, we will contend that the large-$N$ expansion can fail when $N$ is arbitrarily large, even if subleading corrections are seemingly suppressed and the model is well-defined at finite $N$. 

We consider a tensor model \cite{Gurau:2009tw,Gurau:2011aq,Gurau:2011xq,Bonzom:2011zz,Tanasa:2011ur,Bonzom:2012hw,Carrozza:2015adg,Klebanov:2016xxf,Giombi:2018qgp}\footnote{See Ref. \cite{Klebanov:2018fzb} for a review.} generalization of the quantum $p$-spherical model with $p=4$ \cite{Cugliandolo1998Quantum,Cugliandolo1999RealTime,Cugliandolo2001,Anous:2021eqj,Winer:2022ciz},\footnote{See Appendix \ref{quantumspherical} for review.} given by the Hamiltonian
\begin{equation}\label{tensormode}
H=\sum_{\forall j: \ 1 \leq i_{j}\leq N} \frac{(\pi^{i_1 i_2 i_3})^{2}}{2\mu}-\frac{j}{4!}\sum_{\forall j: \ 1\leq i_{j}\leq N}\phi^{i_{1}i_{2}i_{3}}\phi^{i_{1}i_{4}i_{5}}\phi^{i_{6}i_{2}i_{5}}\phi^{i_{6}i_{4}i_{3}}, \quad j=\frac{J}{N^{3/2}} \ .
\end{equation}
$\phi$ and $\pi$ are conjugate position and momentum variables respectively. This model has an $O(N)^{3}$ symmetry. We additionally impose the spherical constraint 
\begin{equation}
\label{psphericcond}
\sum_{\forall j: \ 1 \leq i_{j}\leq N} (\phi^{i_1 i_2 i_3})^{2}=N^{3} \ .
\end{equation}
which is equivalent to imposing that the eigenvalue of the sum of operators on the left is equal to the right-hand side.\footnote{Note that one must project out the kinetic term in Eq. (\ref{tensormode}) that corresponds to Eq. (\ref{psphericcond}). In other words, the Hamiltonian should not have a kinetic term for the radial direction, which is fixed.} The constant $j$ is chosen so that the large-$N$ expansion of the model is well-defined. Notably, the large-$N$ perturbation theory is equivalent to the quantum $p$-spherical model with $p=4$.\footnote{See Refs. \cite{Giombi:2017dtl,gurau2019notestensormodelstensor,Benedetti_2015} for discussions of other bosonic tensor models.} We do not consider the generalization to generic $p$ for simplicity. 

This model is well-defined at any finite, integer $N$. This model is dominated by melonic diagrams at large-$N$ \cite{Klebanov:2016xxf}, which leads to Schwinger-Dyson equations that predict an extensive ground state energy. Nevertheless, we show in Section \ref{sec:thegs} that the ground state energy of this model is superextensive in the number of degrees of freedom, and that, as a consequence, large-$N$ perturbation theory must fail, even when the effective coupling, $\beta J$, is arbitrarily small. 

The failure of perturbative expansions is not a new phenomenon. The theory
\begin{equation}
\mathcal{L}=\int d^{d}x \left [ \frac{1}{2}(\partial \phi)^{2}+\frac{g}{6}\phi^{3} \right ]
\end{equation}
seems fine at any finite order in perturbation theory. Nonetheless, the theory is not well defined non-perturbatively because the potential is unbounded from below. An alternative example is QED, except with imaginary coupling. Dyson argued such a theory would be ill-defined because any physical state would always be unstable to the creation of a large number of particles \cite{Dyson:1952tj}. Notably, neither these models are well defined at finite values of the coupling. In contrast, we emphasize that Eq. (\ref{tensormode}) is a well-defined theory at finite $N$ whose thermodynamics and correlation functions could be calculated on a sufficiently powerful computer. The theory exists; only the large-$N$ expansion is problematic. 


While we are studying a specific model, we think the results have broader implications. Most obviously, the large-$N$ expansion of any model without a non-perturbative justification is somewhat questionable, even if the model is well defined at finite $N$ and weakly coupled. Additionally, our result raises important questions about spin glass systems \cite{Binder1986Spin,Mezard1987,Fischer1991,Nishimori2001,Castellani2005Spin,Mezard2009,Stein2013}. Finally, super extensive systems are interesting in their own right and the tensor model provides a computationally tractable example. These comments are discussed in more detail in Section \ref{sec:conc}. \\

\noindent \textbf{Notation}: In this paper, the index 
$i_{j}$ represents the color-indices of the tensor fields (ranging from $1$ to $N$). All sums over $i_{j}$ are over all values the given index can take. 


\section{Large-$N$ perturbation theory}
\label{sec:theModel}

In this section, we discuss the large-$N$ perturbative expansion of the tensor model.

We consider the partition function 
\begin{equation}
Z[\mathcal{J}^{i_{1}i_{2}i_{3}}]=\int \mathcal{D}\phi^{i_{1}i_{2}i_{3}}\delta(N^{3}-\sum_{i_{j}}(\phi^{i_{1}i_{2}i_{3}})^{2})e^{iS}
\end{equation}
where
\begin{equation}
S=\int dt \sum_{i_{1}i_{2}i_{3}}(\frac{i\mu}{2}(\partial_{t}\phi^{i_{1}i_{2}i_{3}})^{2}+\mathcal{J}^{i_{1}i_{2}i_{3}}\phi^{i_{1}i_{2}i_{3}})+\frac{j}{4!}\sum_{i_{j}}\phi^{i_{1}i_{2}i_{3}}\phi^{i_{1}i_{4}i_{5}}\phi^{i_{6}i_{2}i_{5}}\phi^{i_{6}i_{4}i_{3}} \ .
\end{equation}
The purpose of computing the partition function as a function of the source terms, $\mathcal{J}^{i_{1}i_{2}i_{3}}$, is to compute higher point functions. This function looks very similar to partition functions of more standard tensor models except for the delta-function, which we resolve using an auxiliary field 
\begin{equation}\label{psphericcond2}
\delta(N^{3} -\sum_{i_{j}} 
 (\phi^{i_1,i_2,i_3})^{2})=\int \mathcal{D}z(t)\exp{iz(t)\sum_{i_{j}}(1- (\phi^{i_1,i_2,i_3})^{2} )} \ .
\end{equation}
Therefore, we have a path integral over $N^{3}$ $\phi$ fields and one auxiliary field $z(t)$. Higher point correlators are computed by taking variations of the partition function with respect to the source field. 

We now consider large-$N$ perturbation theory. We can compute the partition function and correlation functions using large-$N$ perturbation theory. In principle, the $z(t)$ terms look problematic. However, since the $z(t)$ field couples to $\mathcal{O}(N^{3})$ fields, the fluctuations are suppressed at large-$N$. The $z(t)$ does gain an expectation value that is $\mathcal{O}(1)$ though, so it contributes in perturbation theory and should be treated as a mass correction to the $\phi$ fields. In this approximation, the free propagator is given by 
\begin{equation}\label{propfree}
\langle T(\phi^{i_{1}i_{2}i_{3}}(t)\phi^{i_{4}i_{5}i_{6}}(t))\rangle|_{\textrm{Free}}=\delta^{i_{1}i_{4}}\delta^{i_{2}i_{5}}\delta^{i_{3}i_{6}}\mathcal{G}_{0}(t,t')
\end{equation}
where
\begin{equation}\label{deffreeprop}
i(\mu \partial_{t}^{2}+z(t))\mathcal{G}_{0}(t,t')=\delta(t-t') \ .
\end{equation}
Perturbation theory will involve diagrams using the propagator in Eq. (\ref{propfree}) with some contour prescription that depends on the time ordering of the observable being computed. 

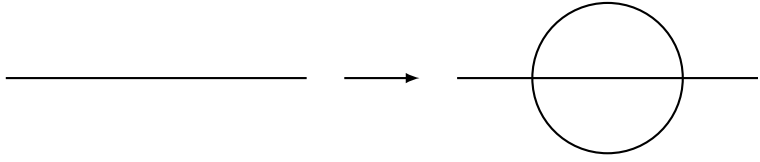
\begin{figure}\center
\begin{tikzpicture}

\draw[thick] (4, 0) -- (8, 0);     
\draw[thick] (6, 0) circle (1);        

\draw[thick, ->] (2.5, 0) -- (3.5, 0);     

\draw[thick] (-2, 0) -- (2, 0);     

\end{tikzpicture}
\caption{The replacement of a single propagator with a melonic sub-diagram.}
\label{melonicrep}
\end{figure}

\begin{figure}\center
\begin{tikzpicture}

\draw[thick] (-2, 0) -- (2, 0);     
\draw[thick] (0, 0) circle (1);        

\draw[thick] (-1.5, 0) -- (-0.5, 0);     

\draw[thick] (4, 0) -- (8, 0);     
\draw[thick] (6, 0) circle (1);        
\draw[thick] (6, 0.5) circle (0.5);     

\draw[thick] (-2, 0) -- (2, 0);     

\node at (3, 0) {Vs.};

\end{tikzpicture}
\caption{The diagram on the left is a melonic diagram with $\mathcal{O}(1)$ scaling at large-$N$. The diagram on the right is a non-melonic diagram with sub-leading scaling at large-$N$ in the tensor model.}
\label{comparegraph}
\end{figure}
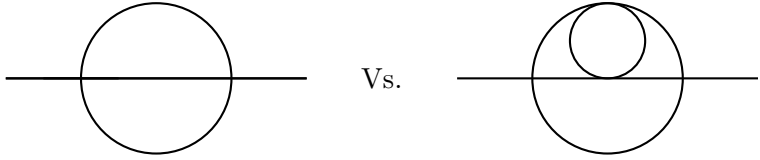

An important simplification at large-$N$ is that melonic diagrams dominate the diagrammatic expansion \cite{Klebanov:2016xxf}. The definition of a melonic diagram is recursive. The propagator is the simplest melonic diagram. To construct a new melonic diagram, take any propagator in the original melonic diagram and perform the replacement in Fig. \ref{melonicrep}. Any melonic diagram can be constructed by some iteration of this procedure. With relatively little effort, one can show that any two-point melonic diagram is $\mathcal{O}(1)$ at large-$N$ using this recursive definition. Consider the first graph in Fig. \ref{comparegraph}. For each line, there are three indices. Each closed cycle of indices gives a factor of $N$. Without the coupling, the diagram therefore scales as $\mathcal{O}(N^{3})$ in the large-$N$ limit because there are three complete cycles. The coupling factor, $j^{2}$, scales as $\mathcal{O}(N^{-3})$, so the diagram is $\mathcal{O}(1)$ in the large-$N$ limit. This result implies the operation in Fig. \ref{melonicrep} does not change the large-$N$ scaling of a graph, which in turn implies that all graphs created by iterations of Fig. \ref{melonicrep} on the propagator scale as $\mathcal{O}(1)$ at large-$N$. The fact that all non-melonic diagrams are sub-leading in $N$ is more non-trivial to prove. An example of a non-melonic diagram is the second graph in Fig. \ref{comparegraph}. We refer the reader to Ref. \cite{Klebanov:2016xxf} for a detailed proof that non-melonic graphs are all sub-leading.\footnote{See also Refs. \cite{Gurau:2009tw,Carrozza:2015adg}.}

With the knowledge that melonic diagrams dominate perturbation theory, we can compute observables like the free energy and the spectrum using the Schwinger-Dyson equations as in Ref. \cite{Klebanov:2016xxf}. We first assume the exact propagator takes the form
\begin{equation}\label{exactprop}
\langle T(\phi^{i_{1}i_{2}i_{3}}(t)\phi^{i_{4}i_{5}i_{6}}(t))\rangle=\delta^{i_{1}i_{4}}\delta^{i_{2}i_{5}}\delta^{i_{3}i_{6}}\mathcal{G}(t,t') 
\end{equation}
at large-$N$ and can be accurately computed using perturbation theory about the free theory. The two-point Schwinger-Dyson equation is then
\begin{equation}\label{twopointSDeq}
\mathcal{G}(t_{1},t_{2})=\mathcal{G}_{0}(t_{1},t_{2})+j^{2}N^{3}\int dt dt' \mathcal{G}_{0}(t_{1},t)\mathcal{G}(t,t')^{3}\mathcal{G}(t',t_{2}) \ .
\end{equation}
By iterating Eq. (\ref{twopointSDeq}), we obtain the exact propagator as the sum of melonic diagrams given by large-$N$ perturbation theory. Eq. (\ref{twopointSDeq}) is a resummation of such diagrams into an integral equation. However, Eq. (\ref{twopointSDeq}) is insufficient to solve for $\mathcal{G}(t,t')$ due to $z(t)$, which is currently unfixed and appears in $\mathcal{G}_{0}(t,t')$. To solve for $z(t)$ and $\mathcal{G}(t,t')$, we need the additional constraint
\begin{equation}\label{auxcon}
\mathcal{G}(t,t)=1 \ .
\end{equation}
To derive Eq. (\ref{auxcon}), take the time-ordered expectation value of the constraint, Eq. (\ref{psphericcond}), and note that the left-hand side becomes the exact propagator using Eq. (\ref{exactprop}). Combining the Schwinger-Dyson equations (\ref{deffreeprop}), (\ref{twopointSDeq}) and (\ref{auxcon}), we could in principle solve for the exact propagator, but this is very difficult. The above procedure can be generalized to higher point correlators, as done for a similar fermionic tensor model in Ref. \cite{Klebanov:2016xxf}.

%
%

Crucially, the Schwinger-Dyson equations predict the energy spectrum scales as $\mathcal{O}(N^{3})$ at finite temperature\footnote{At low temperatures, there could be replica-symmetry breaking. We consider sufficiently high temperatures to prevent this from occurring.} and is therefore extensive in the number of degrees of freedom. To see this without an explicit calculation, note that the Schwinger-Dyson equations (\ref{deffreeprop}), (\ref{twopointSDeq}) and (\ref{auxcon}) are equivalent to the quantum $p$-spherical model reviewed in Appendix \ref{quantumspherical} with $N^{3}$ fields and $p=4$. The large-$N$ limit of the $p$-spherical model is well studied and the reader is referred to Refs. \cite{Anous:2021eqj,Winer:2022ciz} for review. Since the Schwinger-Dyson equations of the $p$-spherical model predict an extensive energy spectrum, so must the tensor models' large-$N$ Schwinger-Dyson equations.

\section{A super-extensive state}
\label{sec:thegs}

In this section, we exhibit a state with super-extensively negative energy, that is a state whose energy is less than $cM=cN^3$ as $M$ becomes large, where $c$ is some negative constant.

We begin our construction by examining the position eigenstate where $\phi^{111}=N^{3/2}$ and all other $\phi^{i_{1}i_{2}i_{3}}=0$. This state satisfies the spherical constraint. The potential energy is
\begin{equation}
    V=-\frac {j}{4!} (\phi^{111})^{4}=-\frac {j}{24}N^6=-\frac {J}{24} N^{4.5}
\end{equation}
Note that this potential energy is super-extensively negative: it scales faster than the number of degrees of freedom $M=N^3$. It also spontaneously breaks the $O(N)^3$ symmetry: there is a huge manifold of other trial states related to it by rotating the three indices.

Unfortunately, the construction of a super-extensively negative state is not quite done. The position eigenstate we proposed has infinite momentum and infinite kinetic energy. To remedy this problem, we could replace our original trail state with a coherent state with a highly localized position for all $\phi^{i_{1}i_{2}i_{3}}$. If we did not have the spherical constraint (\ref{psphericcond}), this would be sufficient to show that the state is superextensive because the kinetic energy of such a state would trivially be extensive and unable to compete with the potential term. Intuitively, the same should be true even with the spherical constraint (\ref{psphericcond}) because, for a sufficiently localized wavefunction, a $d$-sphere looks like $\mathbb{R}^{d}$. 

In principle, quantizing a system with a non-linear constraint is subtle \cite{Batalin:1986fm,Kleinert:1997em}. We will circumvent this issue by quantizing the system without the spherical constraint and then consider states that are highly localized in the radial direction. Consider the following trial wavefunction without the spherical constraint:
\begin{equation}
\Psi_{\textrm{trial}}^{\sigma,\sigma_{r}}=\Theta(\phi^{111})\frac{\phi^{111}}{r}\frac{e^{\frac{-(r-N^{3})^{2}}{4\sigma_{r}^{2}}}}{(2\pi \sigma_{r})^{1/4}}\prod_{(i_{1}i_{2}i_{3})\neq (111)} \frac{1}{(2\pi \sigma^{2})^{1/4}}e^{\frac{-(\phi^{i_{1}i_{2}i_{3}})^{2}}{4\sigma^{2}}} \ ,
\end{equation}
where 
\begin{equation}
r=\sqrt{\sum_{i_{1}i_{2}i_{3}} (\phi^{i_{1}i_{2}i_{3}})^{2} } \ .
\end{equation}
In the limit that $\sigma,\sigma_{r}\rightarrow 0$, this state is normalized to unity. The exponential pre-factor containing $r$ is meant to localize the state to an eigenstate of $r$ that obeys the spherical constraint (\ref{psphericcond}) as $\sigma_{r}\rightarrow 0$. Due to the Heaviside function, the $\phi^{111}$ prefactor is necessary to impose the boundary condition that the wavefunction is zero when $\phi^{111}=0$. The remaining Gaussians localize all the $\phi^{i_{1}i_{2}i_{3}}$ except $\phi^{111}$. We compute the kinetic energy of this state in the limit $1>>\sigma>>\sigma_{r}$ at large-$N$, finding the leading contribution to the kinetic term is
\begin{equation}\label{kineticenergy}
\textrm{Kinetic Energy}\sim \frac{1}{8\mu \sigma_{r}^{2}}+\frac{N^{3}}{8\mu \sigma^{2}}+\ldots \ .
\end{equation}
The first term in Eq. (\ref{kineticenergy}) corresponds to how the contribution to the kinetic energy from the radial direction is divergent for a state with definite radial position; this term can be subtracted without issue. The second term is physical and corresponds to the actual kinetic energy of the highly-localized particle on the $(N^{3}-1)$-sphere. As long as we impose that $\sigma>N^{-3/4}$, this term cannot compete with the potential contribution at asymptotically large-$N$ and the state has a super-extensive negative energy.

We thus have a state with super-extensive negative energy. This does not imply that the identified state is necessarily the exact ground state or even an energy eigenstate. But we now know the Hamiltonian has at least one eigenvalue that is superextensively negative in $M$. This has an enormous impact on the partition function and the free energy. It means that for any finite $\beta$, even those corresponding to fairly high temperatures, the partition function grows as $Z\sim e^{\# \beta M^{1.5}}$, not $Z\sim e^{\# \beta M}$ as the perturbation theory in Section \ref{sec:theModel} predicts. Thus, this perturbative calculation of the free energy and the partition function must be catastrophically wrong, even at fairly weak values of the coupling.

\section{Related Models}
The non-perturbative failure we discuss here is not unique to equation \ref{tensormode} with positive $j$. There is a construction that works just as well with negative $j$. Take any small $N_0$ and find a configuration $\phi_0$ with negative potential energy. Then scale the system up to $N$ by multiplying each element of $\phi$ by $\sqrt{\frac{N^3}{N_0^3}}$, and letting $\phi_{ijk}$ be 0 whenever $i$, $j$, or $k$ is greater than $N_0$. This configuration will satisfy the spherical constraint, having the sum of the squares of its elements go as $N^3$. The energy will go as $N^6\textrm{ (tensor elements)}\times N^{-3/2}(j)=N^{4.5}=M^{1.5}$. Thus, given any negative-energy state, we can construct a state with super-extensively negative energy. This logic works for a huge number of situations, including tensors with symmetric or anti-symmetric constraints, and positive or negative $j$. 

With some adaptation, it even works for systems with spin constraints. In such systems, the spherical constraint has been replaced with the much stronger constraint that each element is $\pm 1$. This gives us a tensor analogue of the $p$-spin model first studied by Gardner \cite{gardner1985}. In such cases, a brute rescaling will obviously violate the constraint on the spins. However, after such a rescaling, one can do a random $O(N)$ rotation. The tensors will still be low-rank, but each element will be $O(1)$. If we round each element to $\pm 1$, the energy will be superextensive, since the new tensor will still have overlap with the low-rank one. However, the new tensor will satisfy the Ising constraint, thus giving us super-extensively negative energy even in an Ising-type model.

Finally, we wish to discuss matrix analogues of this phenomenon. While the tensor models all have the property that they share a large-$N$ expansion with a safe, well-defined disordered systems, to our knowledge the matrix model does not. Nonetheless, one can imagine a matrix model with the spherical constraint $\lambda(\tr M^2-N^2)$,  with interaction term $\frac{g}{24 N}\tr M^4$. Using this normalization the ribbon diagrams all seem to contribute at the same order. However a massively singular matrix with eigenvalues $(N,0,\dots)$ would have potential proportional to $N^3$. If $g$ is negative, this would be enough to dominate the partition function. For positive $g$, there is no such thing as a superextensively negative configuration, since the potential is positive-definite and there are no negative configurations at all.

\section{Discussion}
\label{sec:conc}

We have studied a model for which large-$N$ perturbation theory seems valid, particularly at very weak coupling, but gives a catastrophically wrong result. This result has several implications and future directions. 

We restricted ourselves to a model with a single flavor of $\phi^{i_{1}i_{2}i_{3}}$ fields. However, a model that more closely resembles the construction in Ref. \cite{Witten:2016iux} would have four flavors of scalars, such as 
\begin{equation}\label{tensormode2}
H=\sum_{\substack{1\leq k\leq 4\\
                  \forall j: \ 1 \leq i_{j}\leq N}} \frac{(\pi_{k}^{i_1 i_2 i_3})^{2}}{2\mu}-j\sum_{\substack{
                  \forall j: \ 1 \leq i_{j}\leq N}}\phi_{1}^{i_{1}i_{2}i_{3}}\phi_{2}^{i_{1}i_{4}i_{5}}\phi_{3}^{i_{2}i_{4}i_{6}}\phi_{4}^{i_{3}i_{5}i_{6}}, \quad j=\frac{\sqrt{6}J}{N^{3/2}} \ ,
\end{equation}
with the spherical constraint 
\begin{equation}
\label{psphericcond2}
\sum_{\substack{1\leq k\leq 4\\
                  \forall j: \ 1 \leq i_{j}\leq N}} (\phi_{k}^{i_1 i_2 i_3})^{2}=4N^{3} \ .
\end{equation}
Unlike the model we consider, this model has both super-extensive negative and positive potential energy states. Nonetheless, melonic diagrams also dominate the large-$N$ expansion of this model, so large-$N$ perturbation theory should also (incorrectly) predict an extensive energy spectrum.   

It would be interesting to investigate whether models involving fermions can exhibit a similar breakdown in large-$N$ perturbation theory. For example, Ref. \cite{Witten:2016iux} studied a fermionic tensor model that mimics the behavior of the SYK model, but without disorder. This tensor model could theoretically display a super-extensive ground state, potentially challenging the validity of its large-$N$ perturbation theory. However, the anti-commuting nature of fermions seems to complicate the construction of such a super-extensive state. For instance, Ref. \cite{Klebanov:2018nfp} established bounds on the ground state energy in a different fermionic tensor model, demonstrating that a super-extensive state is not possible. The (non)existence of super-extensive states remains an intriguing question for tensor models.

Our model offers intriguing implications for spin glass systems. Without the super-extensive ground state present in this bosonic tensor model, the equivalence of large-$N$ perturbation theory of the tensor model with the $p$-spherical model suggests the existence of a spin-glass phase at low temperatures. Although our model could not exhibit a true spin-glass phase due to the super-extensive ground state, it raises the possibility that another tensor model could. In such a model, the disorder might arise from varying $N$ rather than through explicitly random couplings. For other proposals of spin-glass systems without explicit disorder, see Refs. \cite{PhysRevLett.85.836,PhysRevB.64.174203,Kamber_2020}.

In addition to broader implications, the tensor $p$-spherical model is interesting in its own right. The model has an $O(N)^3$ symmetry, and the trial ground state we supplied spontaneously breaks this symmetry. It would be interesting to better understand whether this symmetry is in fact broken in the large-$N$ limit.

\subsection*{Acknowledgments}

We thank Razvan Gurau, Igor Klebanov, Juan Maldacena, Erez Urbach and Edward Witten for useful discussions. MW acknowledges DOE grant DE-SC0009988. AH is grateful to the Simons Foundation as well as the Edward and Kiyomi Baird Founders' Circle Member Recognition for their support.

\appendix

\section{The quantum $p$-spherical model}\label{quantumspherical}

The quantum $p$-spherical model \cite{Cugliandolo2001,Biroli_2001,PhysRevB.23.4661} is defined by the Hamiltonian
\begin{equation}\label{phsericalmod}
H=\sum^{N}_{i=1}\frac{\pi_{i}^{2}}{2\mu}+\sum_{(i_{1},i_{2},i_{3},i_{4})} J_{i_{1},i_{2},i_{3},\ldots,i_{p}}\phi_{i_{1}}\phi_{i_{2}}\phi_{i_{3}}\ldots \phi_{i_{p}}
\end{equation}
where $1<i_{j}\leq N$ and the $J_{i_{1},i_{2},i_{3},\ldots,i_{p}}$ are independent Gaussian random variables with mean zero and standard deviation 
\begin{equation}
\mathbb{E}\left [ (J_{i_{1},i_{2},i_{3},i_{4}})^{2} \right ]=\frac{p! J^{2}}{2 N^{p-1}} \ .
\end{equation}
The $\pi_{i}$ and $\phi_{i}$ are conjugate variables. We additionally impose the constraint $\sum_{i}^{N}\phi_{i}^{2}=N$. We now restrict to $p=4$ for simplicity although the strategy applies for general $p$.

The constraint can be implemented in the path integral by including an integral over an auxiliary field $z(t)$
\begin{equation}\label{sum1}
\delta(N-\sum_{i}^{N}\phi_{i}^{2})=\int \mathcal{D}z(t)e^{iz(t)(N-\sum_{i}^{N}\phi_{i}^{2})} \ .
\end{equation}
To derive the large-$N$ limit, one first re-writes the action as a function of 
\begin{equation}\label{identify}
\mathcal{G}(t,t')=\frac{1}{N}\sum \phi_{i}(t)\phi_{i}(t')
\end{equation}
and adds auxiliary fields $\mathcal{F}(t,t')$ to impose (\ref{identify}). One then performs an ensemble average over the couplings. See Section 2 of Ref. \cite{Winer:2022ciz} for details. The end result is the Dyson-Schwinger equations 

\begin{equation}
\begin{split}
&i(\mu\partial_{t}^{2}+z)\mathcal{G}(t,t')+\int dt'' \mathcal{F}(t,t'')\mathcal{G}(t'',t')=\delta(t-t') \ , \\
&\mathcal{F}(t,t')=J^{2}\mathcal{G}(t,t')^{3}, \quad \mathcal{G}(t,t)=1 \ .
\end{split}
\end{equation}
Solving for $\mathcal{F}(t,t')$, one finds the $4$-spherical model Schwinger-Dyson equations are equivalent to those of the tensor model, Eqs. (\ref{twopointSDeq}) and (\ref{auxcon}), at large-$N$.

At sufficiently low temperatures, the variance of the partition function under variations of the random couplings is parametrically large, which is a signature of the spin-glass phase. However, this phenomena can actually be studied in perturbation theory by computing the replica-symmetry breaking saddle at large-$N$. 

\section{The physics of the ground state manifold}\label{groundstateman}
The state discussed in Section \ref{sec:thegs} does not obey Eq. (\ref{exactprop}). By giving vacuum expectation values to the fields, we turn our quartic interaction into a host of cubic and quadratic terms. These quadratic terms, in particular, vastly change the propagator. They include cross-terms allowing, for instance, the field $\phi_{123}$ to become the field $\phi_{145}$. Thus any description would need to involve a far greater number of propagators than considered in Section \ref{sec:theModel}. There is, however, a compact description of the low energy physics in terms of Goldstone modes.

The state exhibited in Section \ref{sec:thegs} is actually part of a larger manifold parameterized by three $N-$dimensional vectors $v^{i_1}_1,v^{i_2}_{2},v^{i_3}_{3}$, each with magnitude $\sum_a v^{a2}=N$. Given these parameters we can construct a position eigenstate $\phi^{i_1i_2i_3}=v_{1}^{i_1}v_{2}^{i_2}v_{3}^{i_3}$.  This collection of states breaks the $O(N)^3$ symmetry of the Hamiltonian to $O(N-1)^3$.

Let us write an action in terms of these $3N$ variables. The potential is a constant, $-N^{4.5}$, over the full manifold. The kinetic term, which can be written $\frac \mu 2 \sum_{i_1i_2i_3}\left(\partial_{t} \phi^{i_1i_2i_3}\right)^2$. This can be expressed in terms of the $v$s as 
\begin{equation}KE=\frac {N^2\mu} 2 \sum_{k,i}\left(\partial_{t}v_{k}^{i_k}\right)^2.\end{equation}
The effective mass of these modes grows with $N$, meaning that at large-$N$ one can introduce angular momenta around these spheres for basically no energy penalty. In other words, the symmetry is spontaneously broken and the model develops Goldstone modes.

\bibliographystyle{apsrev4-1long}
\bibliography{GeneralBibliography.bib}
\end{document}